\documentclass[aps,prb,reprint,groupedaddress]{revtex4-1}

\everymath={\fam0 }
\usepackage{threeparttable}
\usepackage{graphicx}

\begin{document}


\title{M\"ossbauer parameters of  $^{57}$Fe substituents in the topological insulator $\bf Bi_2Se_3$}


\author{Jorge H. Rodriguez}
\email[]{jhrodrig@purdue.edu}
\affiliation{Department of Physics and Astronomy, Purdue University, West Lafayette, IN 47907, USA}



\begin{abstract}
	$^{57}$Fe M\"ossbauer spectroscopy can probe several  \emph{local} structural, electronic and magnetic properties
	of Fe-containing systems. However, to establish a direct relationship between these properties and a system's
	geometric structure, the experimental M\"ossbauer parameters need to be analyzed \emph{via} electronic structure
	calculations.
	Herein, structural, electronic and magnetic effects of  iron substituents in the topological insulator $Bi_2Se_3$,
	as uniquely probed by $^{57}$Fe	M\"ossbauer spectroscopy,
	have been determined   \emph{via} spin-polarized electronic structure calculations.
	The iron ion substituents, of \emph{nominal} Fe$^{3+}(S = 5/2)$ oxidation and spin state,
	are unequivocally
	shown to substitute  Bi$^{3+}$ sites in epitaxial $Bi_2Se_3$ thin-films used for M\"ossbauer measurements.
	Concomitant with iron substitution, \emph{localized}
	structural rearrangements take place whereby the longer Bi-Se bonds of the native system
	are replaced by significantly
	shorter Fe-Se counterparts in the Fe-containing system.
	The resulting distorted-octahedral environment about substituent iron ions
	gives rise to characteristic M\"ossbauer parameters 	($\delta_{Fe} \approx$  0.51 mm/s, $\Delta E_{Q} \approx$  0.20 mm/s)
	which have been calculated  
	in excellent  agreement with
	measured values for Fe-doped  $Bi_2Se_3$ thin films. 
	Consistent with a substituent Fe$^{3+}$ ion's \emph{nominal}
	high-spin electronic configuration ($t_{2g}^{\uparrow \uparrow \uparrow} e_g^{\uparrow \uparrow}$),
	an Fe-centered  spin density                       
	has been established which, nevertheless,                      
	extends towards neighboring Se atoms \emph{via} direct Fe-Se bonding and concomitant Fe(d)-Se(p) hybridization.  
\end{abstract}

\pacs{}

\maketitle

\section{INTRODUCTION}


%

\begin{figure}[h!]
	  \begin{tabular}{c}
		  \multicolumn{1}{l}{\includegraphics[width=0.500\textwidth]{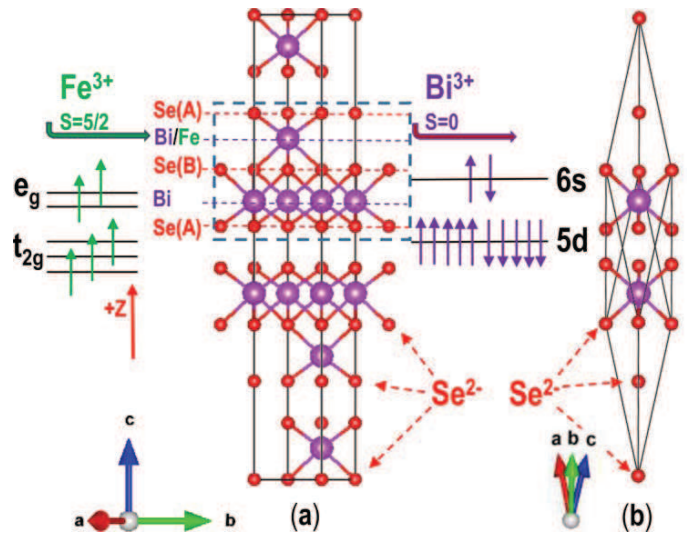}} \\
\end{tabular}
\vskip -9pt
\caption{\label{Fig01} (a) Unit cell  of  $Bi_2Se_3$ displaying  Bi$^{3+}$ (purple) and Se$^{2-}$ (red) ions. 
	The three  \emph{quintuple} layers, stacked along the  z axis, are displayed.
	The \emph{local} environment of each Bi atom,  in the presence of six chemically-bound Se atoms, 
	is \emph{distorted octahedral}. 
	A magnetic Fe$^{3+}$(S=5/2) ion  substitutes a non-magnetic Bi$^{3+}$(S=0) ion within a single \emph{quintuple} layer which is enclosed in a dashed box.    
(b) Primitve cell displaying the Se-rich \emph{distorted octahedral}  environment of each Bi atom.}
\end{figure}

Dibismuth triselenide ($Bi_2Se_3$), a three-dimensional topological insulator (TI) and thermoelectric material,
has a unit cell  characteristically
composed of three \emph{quintuple} layers stacked along the c (z) axis (Fig.~\ref{Fig01}).
Intra-\emph{quintuple} Bi-Se interactions are dominated by strong chemical bonding whereas inter-\emph{quintuple}
interactions are due to weaker van der Waal forces.~\cite{Seiko-1963,Wyckoff1964,Heremans-2017} 
Whereas the Bi atoms of each \emph{quintuple} are equivalent, there are two structurally-inequivalent Se layers whose atoms
are herein denoted  as Se(A) and Se(B).
Some transition metal (TM) ions such as Fe$^{3+}$ can act as Bi$^{3+}$ substituents  and, in addition
to their net positive charge, carry a number of spin-unpaired  electrons.
Fe$^{3+}$ ions in their high-spin configuration                        
nominally carry five unpaired
3d-shell electrons leading to a sizable  spin ($S_{Fe} = 5/2$) and  correspondingly high magnetic
moment ($\vec{\mu}_{Fe}=-g_s\mu_B\vec{S}_{Fe}/\hbar$).
Interaction of the atomic orbitals of a substituent iron ion with those of
neighboring, chemically-bound, Se atoms leads to some degree of Fe(3d) electron delocalization.
This effect  changes the \emph{net} charge and magnetic moment of Fe substituents,                          
relative to their \emph{nominal} values, to
a degree dependent on the host
material's \emph{immediate} electronic and structural  environments.

Two parameters characteristic of an iron ion's  electronic configuration and  chemical-structural
environment are directly  measured by $^{57}Fe$ M\"ossbauer spectroscopy, namely          
isomer shifts ($\delta_{Fe}$) and  quadrupole splittings ($\Delta E_{Q}$).                                      
Spin-polarized electronic structure (SPES) calculations have been used to accurately predict structure-dependent $^{57}Fe$
M\"ossbauer parameters~\cite{Rodriguez2013-HI,Chachiyo-2012} while, at the same time, establishing correlations with other properties of
iron ions  such as their  \emph{net} charges, 
magnetic moments, and spin densities.~\cite{Wang2014-PRB}
Thus, by means of spin-polarized calculations
one can  establish 
correlations between M\"ossbauer parameters and concomitant                                   
electronic, magnetic and structural reorganizations of the host  $Bi_2Se_3$  material 
caused by  iron  substituents.

\begin{figure}[h]
	  \begin{tabular}{cc}
		  \multicolumn{2}{c}{\includegraphics[width=0.475\textwidth]{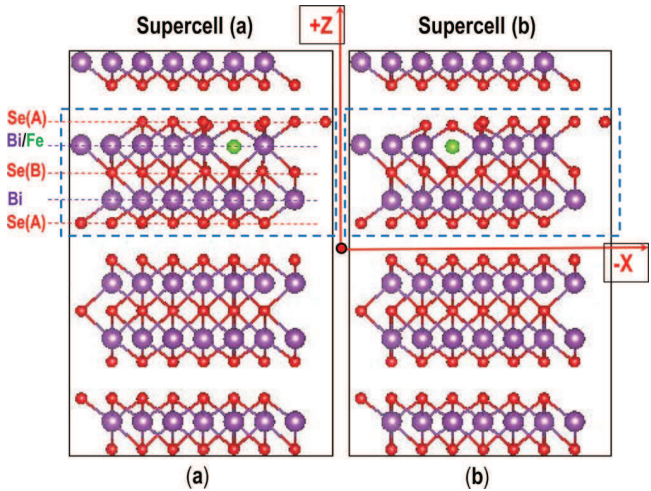}}    
\end{tabular}
\vskip -9pt
\caption{\label{Fig02} Two structurally optimized $3 \times 3 \times  1$ $Bi_2Se_3$ supercells,
	(a) and (b), each containing a
magnetic iron substituent (green). Bi and Se atoms are shown in purple and red, respectively.
The \emph{quintuple} in each supercell containing the  magnetic substituent is shown in a dashed box.}
\end{figure}

\begin{figure}[h!]
	  \begin{tabular}{c|c}
		  {\bf(a)} & {\bf(b)}  \\
		  {\bf Unsubstituted Fixed-Cell} & {\bf Unsubstituted Fixed-Cell} \\
		  \includegraphics[width=0.225\textwidth]{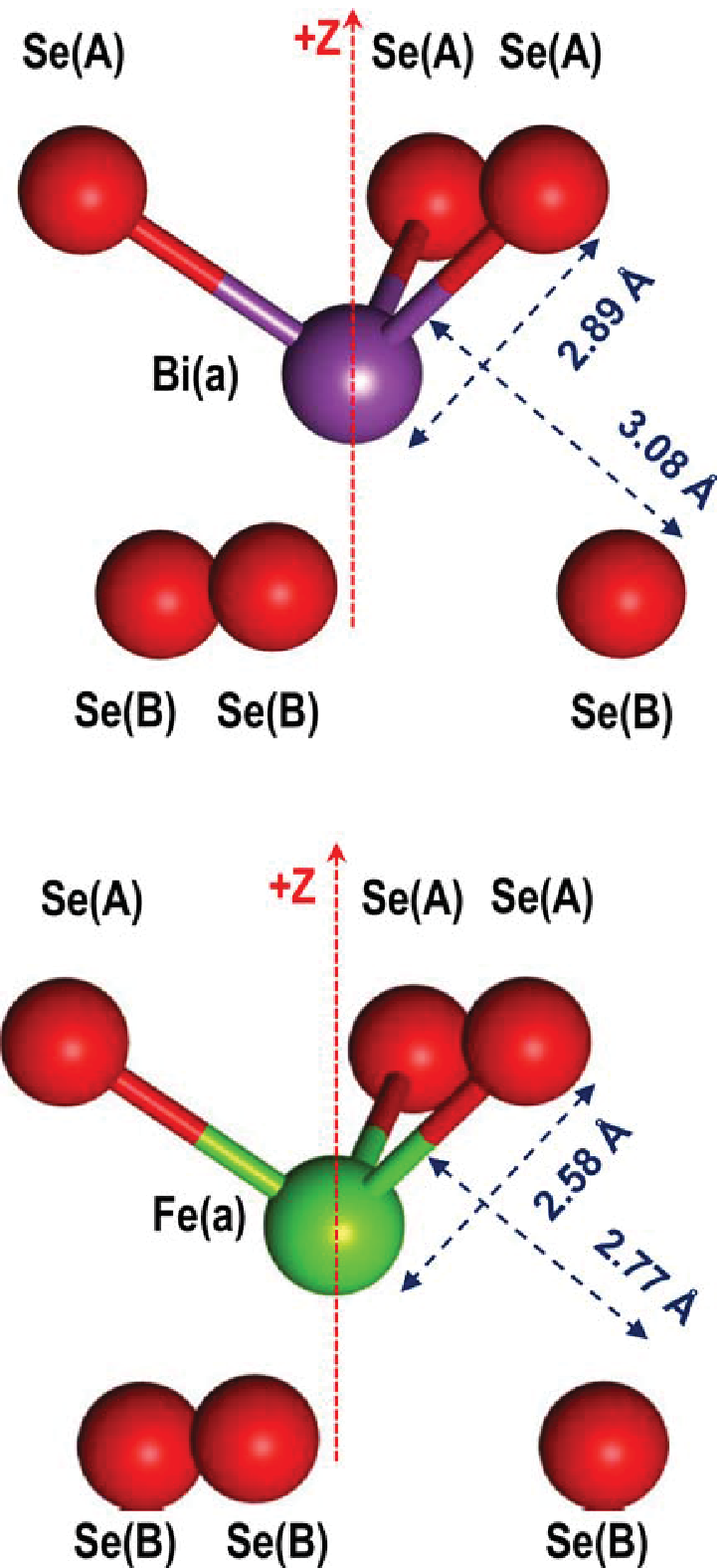}  &
		  \includegraphics[width=0.205\textwidth]{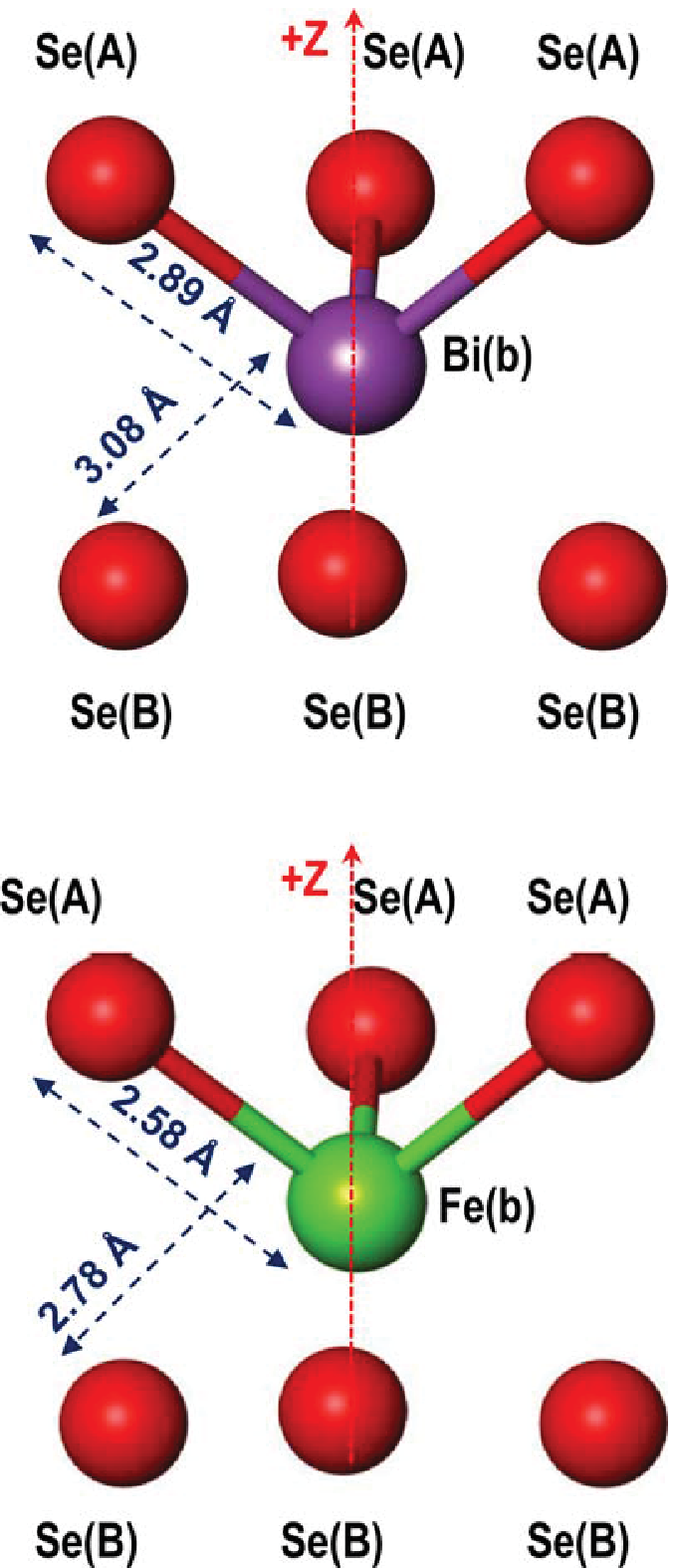}  \\
                   {\bf Substituted Fixed-Cell} &  {\bf Substituted Fixed-Cell} \\
		  {\bf(c)} & {\bf(d)}  \\
			   &           \\
		  \multicolumn{2}{c}{{\bf $^{57}$Fe M\"ossbauer Spectra}} \\ 
		  \includegraphics[width=0.215\textwidth]{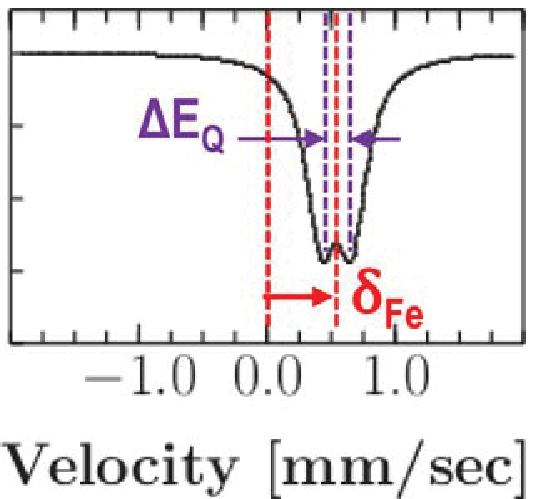}  &
		  \includegraphics[width=0.215\textwidth]{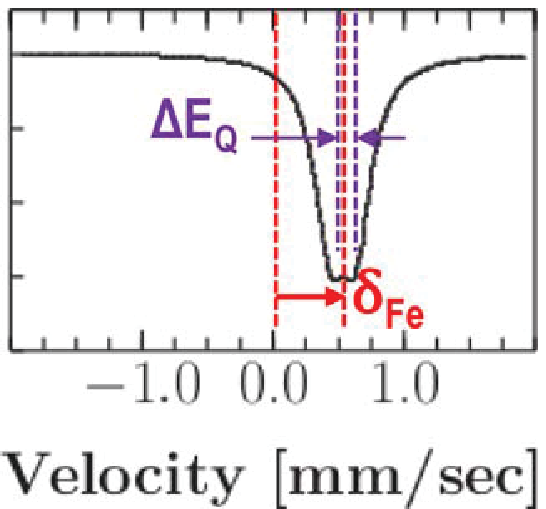}  \\
                 {\bf(e)} & {\bf(f)}  \\
	  \end{tabular}
	  \vskip -6pt
	  \caption{\label{Fig03} (a,b) Se-rich local structural environments of Bi atoms in native $Bi_2Se_3$ 
	     and (c,d) Fe atom  substituents in  $Bi_2Se_3$.
             Significantly different Bi-Se and Fe-Se bond lengths, corresponding
	     to their respective optimized supercell structures, are shown.
	     (e,f) Predicted  absorption M\"ossbauer spectra for supercells (a) and (b), respectively,
	     with $\delta_{Fe}$ and $\Delta E_Q$ values given in Table~\ref{Table02} and
	     calculated for bond lengths and geometries given in (c)-(d).                         
	     Predicted and experimental M\"ossbauer parameters are, within experimental accuracy, in agreement (Table~\ref{Table02}).
	     Each predicted spectrum 
             corresponds to two Lorentzian curves each of 0.28 mm/s width. For each spectrum the separation between the peaks of the two Lorentzians is the
             quadrupole splitting $\Delta E_Q$.}                 
 \end{figure}

Although some experimental techniques probe magnetic properties of a  bulk sample,  
$^{57}Fe$ M\"ossbauer spectroscopy is a powerful technique  to study \emph{local}  electronic,
magnetic and  structural properties of
Fe substituents and their \emph{immediate} structural environment.
Fe-doped epitaxial $Bi_2Se_3$ thin films and their  conversion electron M\"ossbauer spectroscopy
were  reported by Pelenovich and coworkers~\cite{Pelenovich-2015-TSF}   
which provide raw values of the sample's isomer shift ($\delta_{Fe}$) and quadrupole splitting ($\Delta E_{Q}$).
The latter parameters are generally 
highly dependent on the Fe-ion's immediate electronic-structural environment.
More specifically,  $\delta_{Fe}$ and $\Delta E_{Q}$
are sensitive to the structure-dependent 3d-shell electron configuration (e.g. t$_{2g}^{\uparrow \uparrow \uparrow}$ e$_g^{\uparrow \uparrow}$) 
and  spherical or non-spherical symmetry  of the combined electrostatic potential, at the site of the Fe nucleus, generated by  valence 
electrons and/or  surrounding crystalline lattice.

Doping with magnetic TM ions and/or their substitution
for suitable counterparts in  tetradymite topological insulators
has been reported~\cite{KULBACHINSKII-2003,Dyck-2002,Pelenovich-2015-TSF}
and related samples studied by several techniques such as                                         
angle-resolved photoemission spectroscopy (ARPES),
scanning tunneling microscopy (STM) and energy dispersive
X-ray analysis (EDX).~\cite{Yang-2013-PRL,Hsieh-2008,Xu-2012-NAT,Chen-2010-SCI,Li-2013-ML} 
Many studies have focused on the effects of TM doping on TI
surface~\cite{Chen-2010-SCI,Liu-2009-PRL,Heremans-2017} and/or bulk states.~\cite{Yang-2013-PRL,Rosenberg-2012-PRB}                    
In particular, possible effects of the electric and magnetic properties  of iron adatoms
on topologically protected  $Bi_2Se_3$ surface states                         
have been proposed.~\cite{Chen-2010-SCI,Wray-2010-NP,Honolka-2012-PRL,Scholz-2012-PRL}
However, the availability of conversion electron M\"ossbauer spectroscopy data~\cite{Pelenovich-2015-TSF} and its
possible interpretation
by means of \emph{ab initio} SPES~\cite{Rodriguez2013-HI} allows the detailed investigation of other substitutional effects
such as \emph{local} Fe-centered structural changes in host $Bi_2Se_3$  thin films  and
the emergence of partially delocalized spin densities about  Fe substituents.                          
The latter effects can be directly probed  by SPES                            
calculations and, indirectly, 
by  $^{57}$Fe M\"ossbauer spectroscopy. Herein we use SPES to accurately predict $^{57}$Fe M\"ossbauer parameters and show that,
upon  Fe$^{3+}$ for  Bi$^{3+}$ substitution, 
isomer shifts and quadrupole splittings of Fe-doped $Bi_2Se_3$ thin films are directly dependent on the 
structural reorganization of neighboring Se atoms (Figs.~\ref{Fig01}-\ref{Fig03})
and, in addition, that an Fe-centered spin density emerges which propagates onto neighboring Se atoms
(Figs.~\ref{Fig04}-\ref{Fig05}).

Upon  Fe for Bi substitution (Fe$\rightarrow$Bi) we have determined a significant Fe-centered structural  
reorganization of the native $Bi_2Se_3$ host in
qualitative agreement with previous studies.~\cite{Larson-2008-PRB,Zhang-2013-PRB}
A fairly  \emph{distorted octahedral} geometry about the
Fe ions was  found  which is necessary to reproduce  measured M\"ossbauer parameters
for Fe-doped thin films.\cite{Pelenovich-2015-TSF}
Strong Se-atom  reorganization
around the Fe substituents occurs whereby
the former partially \emph{collapse} onto the latter.
Contrary to a nearly spherically symmetric spin distribution expected for  Fe$^{3+}$(S=5/2) in ionic environments,
a cube-like Fe-centered  spin density  was found                                                 
which nevertheless propagates onto neighboring Se atoms.

\noindent
\section{Theory and Methods}
Spin-polarized gradient-corrected  density functional theory               
was used to perform structural relaxations for both types of  $Bi_2Se_3$ structures,
native and Fe-containing  and, thereafter, generate
electron densities needed to evaluate M\"ossbauer  parameters.
Two independent sets of  calculations were performed on two different $3 \times 3 \times  1$ supercells,
labelled (a) and (b),
whereby a different Bi site 
was substituted in each  supercell
by a corresponding  Fe atom as shown in Fig.~\ref{Fig02}.
Both substituted supercell structures were 
relaxed using identical SPES protocols giving rise to two, essentially 
equivalent, geometrically-distorted selenium-rich iron
environments as shown in Figs.~\ref{Fig03}(c,d).
This is consistent with  all Bi atoms, such as Bi(a) and Bi(b), in the unsubstituted supercells being structurally equivalent
to each other and, therefore,
their  respective geometrically-optimized iron substituents, Fe(a) and Fe(b), also being  equivalent to
each other. Each $3 \times 3 \times 1$ supercell included 135 atoms.

For each supercell two types of structural relaxations were done: fixed-cell and variable-cell.  The former kept lattice parameters frozen    
while optimizing  intracell atomic positions  whereas  the latter relaxed both, lattice parameters and  atomic positions.          
All structural optimizations  used  the Perdew-Burke-Ernzerhof exchange-correlation
functional.~\cite{Perdew1996}  
Fixed-cell optimizations used localized orbital basis of DZP quality~\cite{Soler-2002}
with  norm-conserving relativistically-corrected  pseudopotentials as reported by Rivero \emph{et al.}~\cite{Rivero-2015}
Variable-cell  relaxations used the basis  and relativistic effective core potentials 
of Stevens \emph{et al.}~\cite{Stevens1992}.

M\"ossbauer parameters are sensitive to structural variations.
To compare calculated values of $\delta_{Fe}$ and $|\Delta E_Q|$ against experiment and
to study their dependence on various geometric environments,                                       
four types of Fe-containing structures were used for each of the two supercells:                                                   
\begin{table}[h!]   
	\begin{tabular}{ll}
	I.   & 	 Substituted Fixed-Cell                            \\
	II.  & 	 Substituted Variable-Cell \\
	III. &  Unsubstituted Fixed-Cell [Bi$\rightarrow$Fe]    \\
	IV.  &  Unsubstituted Variable-Cell [Bi$\rightarrow$Fe]     

\end{tabular}  
 \end{table}

Structures I and II correspond 
to the dimensions obtained for Fe-substituted supercells \emph{via}  fixed-cell [Figs.~\ref{Fig03}(c,d)] and variable-cell relaxations,
respectively. Structures 
III and IV correspond to the dimensions obtained for unsubstituted supercells \emph{via}  fixed-cell [Figs.~S1(c,d)]\cite{SuppPRB}
and variable-cell relaxations,
respectively, with subsequent replacement of Fe for Bi atoms: [Bi(a)$\rightarrow$Fe(a)] and  [Bi(b)$\rightarrow$Fe(b)].
Supp. Fig.~S1\cite{SuppPRB} illustrates structures III-IV and
Supp. Table~S1\cite{SuppPRB} lists some metric parameters for all, I-IV, structures.

\begin{figure}[h!]
	  \begin{tabular}{c}
		  {\includegraphics[width=0.350\textwidth]{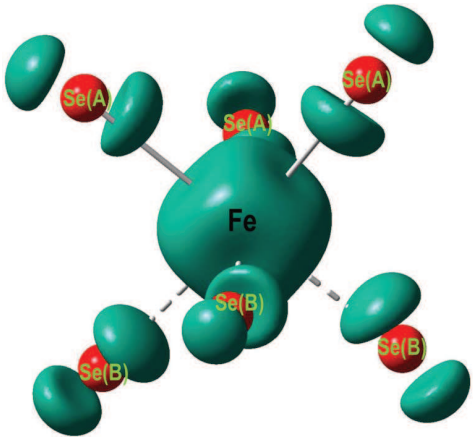}}    \\
 \end{tabular}
 \caption{\label{Fig04} Three-dimensional  plot (in green) of the Fe-centered spin density,
 $\rho_s({\bf r})$,  partially delocalized towards the neighboring Se(A) and Se(B) atoms. The green region
 corresponds to a greater acummulation of majority (spin up), relative to minority (spin down), electrons.}
  \end{figure}

\begin{figure}[h!]
	  \begin{tabular}{|c|c|}
		  \includegraphics[width=0.175\textwidth]{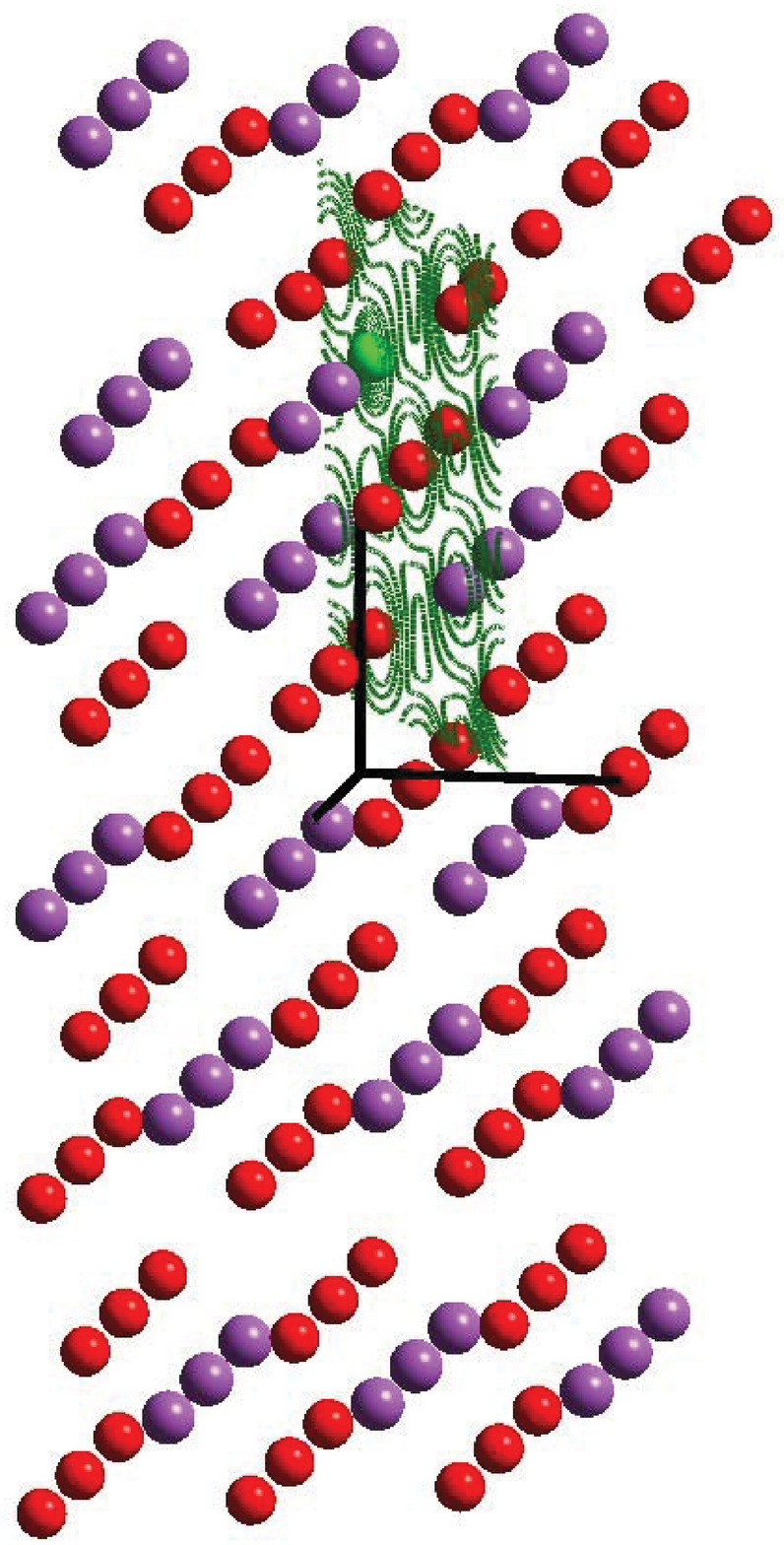} &
                  \includegraphics[width=0.175\textwidth]{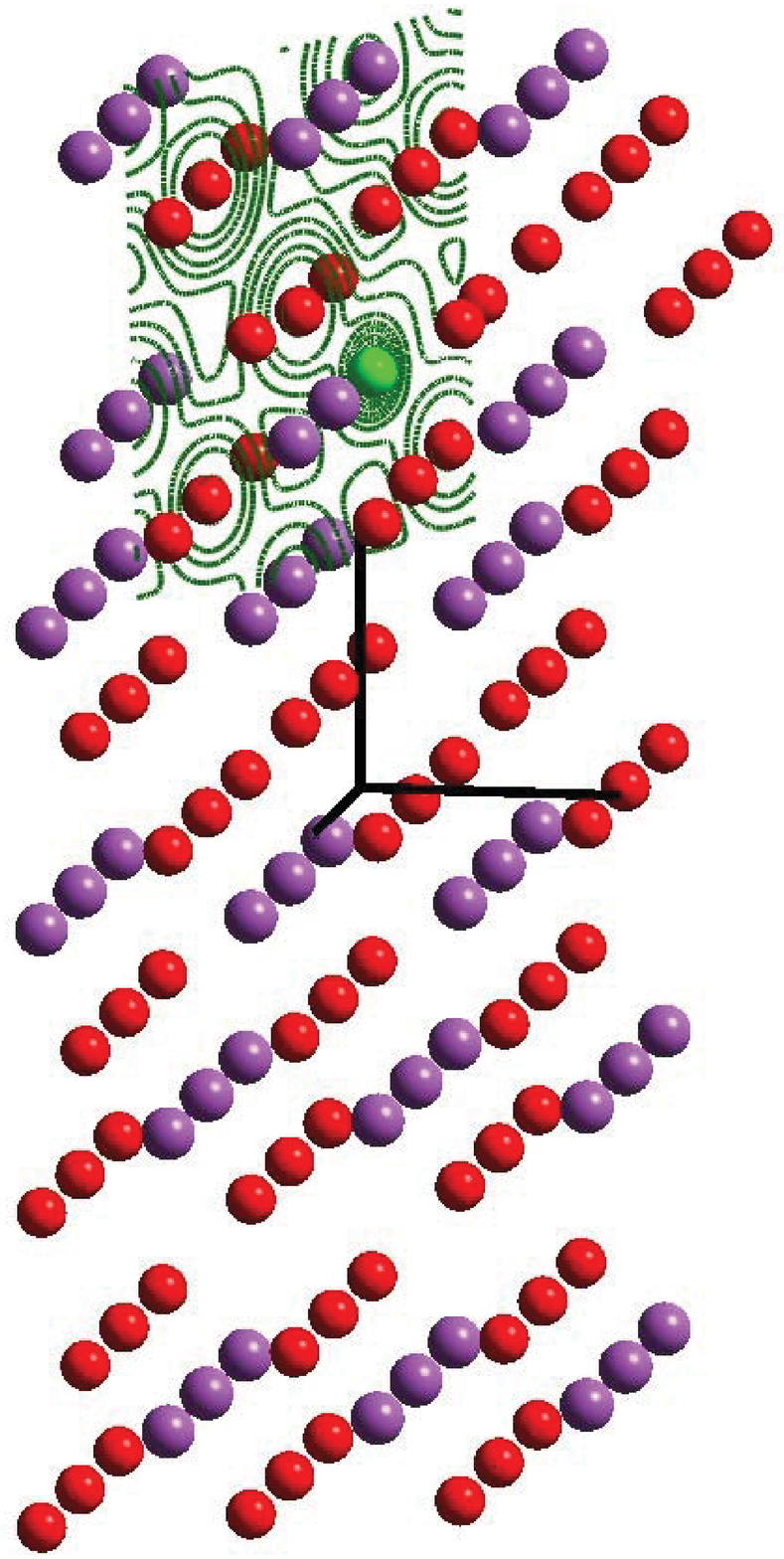} \\
		  {\bf(a)} & {\bf(b)} \\
 \end{tabular}
 \caption{\label{Fig05} Two-dimensional contour plots (in green) of the spin density in quasi-perpendicular planes
	                defined by a) Se(B)-Fe(A)-Se(A) and b) Se(A)-Fe(A)-Se(B).
	               Isosurface value for contours set to 0.008.}
  \end{figure}

Prediction of M\"ossbauer parameters requires evaluation of electron densities at the site of the iron nucleus.
Some authors have used \emph{augmented} plane wave density functional methods      
to evaluate electron densities at the nucleus.~\cite{Zwanziger-2009,Blaha-2010}
Herein, following pseudopotential-based supercell optimization, further all-electron calculations with
Gaussian-type basis on Fe-centered clusters
allowed evaluation of electron densities at the origin. By using a locally developed M\"ossbauer program~\cite{Rodriguez2013-HI}
prediction  of M\"ossbauer  parameters was achieved. Knowledge of ground state
electron densities, $\rho_o({\bf r})$, in turn allows evaluation of ground state energies  ($E_o$) as  functionals of the density:
\begin{eqnarray}
	\rho_o({\bf r}) &=& \sum_i|\phi_i^{\uparrow}({\bf r})|^2 + \sum_j|\phi_j^{\downarrow}({\bf r})|^2 \label{dft}  \\ 
	\label{rho}
	E_o           &=& E_o[\rho_o({\bf r})] \label{E}
\end{eqnarray}
Here, the sets  $\{\phi_i^{\uparrow}({\bf r})\}$ and $\{\phi_j^{\downarrow}({\bf r})\}$
correspond to  occupied
Kohn-Sham orbitals hosting spin-up and spin-down electrons, respectively.
A \emph{formally} Fe$^{3+}$ substituent has a majority of spin-up unpaired 3d-shell electrons
which results in the number of spin-up and spin-down occupied Kohn-Sham orbitals being different. This fact, in turn,
determines the overall system's net spin state and, for Fe-doped $Bi_2Se_3$,
mainly reflects the presence of an Fe-centered magnetic moment.
Thus, the system's spin density, $\rho_s({\bf r})$, as defined in Eq.~\ref{sd} 
is also expected to be Fe-centered in agreement with the numerical results presented in Fig.~\ref{Fig04}:
\begin{equation}
\rho_s({\bf r}) = \sum_i|\phi_i^{\uparrow}({\bf r})|^2 - \sum_j|\phi_j^{\downarrow}({\bf r})|^2 \label{dft}
\label{sd}
\end{equation}
The physical origin of the  M\"ossbauer parameters can be understood in terms of Eq.~(\ref{is}) for the isomer shift
and Eq.~(\ref{eq}) for the quadrupole splitting:
{\small
\begin{eqnarray}
	\delta_{Fe}&=&{2\pi \over 3}Ze^2(\langle R^2\rangle^*-\langle R^2\rangle)\{\rho_{_{Ab}}(r=0)-\rho_{_{So}}(r=0)\} \label{is} \\
        \Delta E_Q&=&{1\over 2}eQV_{zz}(1 + \eta^2/3)^{1\over 2} 	\label{eq} \\
        \eta&\equiv&(V_{xx} - V_{yy})/ V_{zz}  \label{eta}
\end{eqnarray} }
In a M\"ossbauer experiment a moving $^{57}Co$ radioactive source (So) emits a Doppler-shifted
14.4 KeV $\gamma$-ray that excites the nucleus of an absorbing (Ab) Fe ion.
In Eq. (\ref{is}) the difference $-e\{\rho_{_{So}}(r=0) - \rho_{_{Ab}}(r=0)\}$, corresponding to electron
charge densities at the sites of the
emitting source and absorbing nuclei, interacts with the nuclear charge Ze.~\cite{Rodriguez2013-HI}
The term $(\langle R^2\rangle^*-\langle R^2\rangle)$ is the difference of mean square charge radii corresponding
to nuclear excited and ground states, respectively.~\cite{Wertheim-1966-IAEA} In a series of  $^{57}$Fe M\"ossbauer experiments
this nuclear radii difference and  $\rho_{So}(r=0)$ are regarded as  constants and only the term  $\rho_{Ab}(r=0)$ varies for different  samples.

The SPES calculations  evaluate  characteristic electron densities at the nuclei of substituent Fe ions,
$\rho_{{_{Ab=Fe}}}(r=0)$,  in the Se-rich environment of the 
$Bi_2Se_3$ host. Knowledge of electron densities  at iron nuclei
allowed  computation of M\"ossbauer isomer shifts which, due to electron density -  nuclear charge interactions,  
are sensitive to the Fe ion's  electronic charge distribution, magnetic moment  and immediate structural-chemical environment.
Importantly, M\"ossbauer calculations were performed for geometric parameters corresponding
to both, before and after, Fe-induced 
reorganization of $Bi_2Se_3$.
M\"ossbauer parameters were significantly different for relaxed Fe-substituted and
relaxed Fe-unsubstituted supercell  structures. 
Predicted isomer shifts                         
were  compared with  experimentally measured values              
for Fe-doped $Bi_2Se_3$ thin films.
Similarly, by evaluating the components of the electric field gradient (EFG) tensor
at the Fe nuclei  ($V_{xx}$, $V_{yy}$, $V_{zz}$),
which are sensitive to the asymmetric positions of Se$^{2-}$(A) \emph{versus} Se$^{2-}$(B) ions,
we calculated  electric quadrupole splittings ($\Delta E_Q$) which were also compared against
experiment.

\noindent
\section{Results and Discussion}
Both computed parameters, $\delta_{Fe}$ and $\Delta E_{Q}$, were  in excellent agreement with
experiment  but only \emph{after} Fe-induced 
geometric reorganization of the $Bi_2Se_3$ host was taken into account.
This corresponds to the ``Substituted Fixed-Cell"   optimizations 
illustraded by Figs.~\ref{Fig03}(c,d). Due to their slightly different geometric parameters   
``Substituted Variable-Cell" structures displayed close but slightly different M\"ossbauer values (Supp. Table~S1\cite{SuppPRB}).
By contrast, M\"ossbauer values   associated with the longer Fe-Se bond lengths of all ``Unsubstituted  (Bi$\rightarrow$Fe)"  
structures were significantly different from  experiment.
In summary, parameters                               
obtained from all ``Substituted" fixed-cell and variable-cell optimizations
displayed similar trends and were in qualitative agreement with each other. Reults  obtained
for all ``Unsubstituted (Bi$\rightarrow$Fe)" structures were  consistent
within their own group. This indicates that 
it is the presence or absence of Fe-centered reorganization corresponding to the  former   or latter structures, respectively,
that results in M\"ossbauer               
observables in agreement or disagreement with experiment.
Table~\ref{Table02} shows results for ``Substituted Fixed-Cell" relaxed  structures and, for comparison, Supp. Table~S1\cite{SuppPRB} 
includes all other results.

\subsection{Structural Effects of Fe Substitution}
Table~\ref{Table01} displays lattice parameters of the  $Bi_2Se_3$ unit cell~\cite{Seiko-1963,Wyckoff1964}
where the hexagonal structure of the undoped system imposes the  condition   $a = b \neq c$.
Upon substitution  of an Fe ion in lieu of a Bi ion, followed by structural optimization without any
symmetry constraints, the overall hexagonal condition was essentially
maintained.  The relaxed (i.e. energy-minimized)  
Fe-containing 3$\times$3$\times$1  supercells displayed a nearly negligible
difference between the parameters a and b  which are perpendicular to the c axis.            
This minimal asymmetry may reflect the finite size, despite being fairly large,
of the $3 \times 3 \times 1$ supercells whose
structures were optimized without symmetry constraints to ensure that substituent Fe ions
occupy their most energetically-favorable locations.
Despite the overall Fe-unsubstituted  and Fe-substituted  lattice parameters remaining nearly identical,
consistent with EXD measurements~\cite{Li-2013-ML},
\emph{localized} intra-cell structural variations  about the Fe ions were much more noticeable.
As shown in  Fig.~\ref{Fig03}(c,d),   relaxed interatomic distances for the Fe-containing system
[Fe-Se(A): 2.58$\AA$, Fe-Se(B): 2.77$\AA$]                                        
were substantially shorter than                                 
corresponding lengths [Bi-Se(A): 2.89$\AA$,  Bi-Se(B): 3.08$\AA$]                                          
of the relaxed, but unsubstituted, $Bi_2Se_3$ structure.

The present results are qualitatively consistent 
with previous structural relaxation studies  that found neighbooring  Se atoms moving towards the TM ions, with   more distant atoms displaying
much smaller structural variations, ~\cite{Larson-2008-PRB}
and significant displacement towards chromium or iron atoms                      
resulting in shorter Se-Cr or Se-Fe bond lengths, respectively.~\cite{Zhang-2013-PRB}

\begin{table}[h]
	\caption{\label{Table2} Structural parameters of native (unsubstituted) and Fe-containing (substituted) $Bi_2Se_3$ unit cells.} 
	\begin{tabular}{lcccccc} 
		\hline
		        & \multicolumn{6}{c}{Lattice Constants} \\ 
			       & a [$\AA$]  & b [$\AA$]           & c [$\AA$]   & $\alpha$ $[deg]$      & $\beta$ $[deg]$           & $\gamma$ $[deg]$       \\
		\hline
	Native  (Exp.)\cite{Seiko-1963}         & 4.143      & 4.143               & 28.636      & 90.000        & 90.000            &  120.000       \\
	Native  (Exp.)\cite{Wyckoff1964}        & 4.138      & 4.138               & 28.64       & 90.000        & 90.000            &  120.000       \\
	Native\tablenotemark[1]                 & 4.234      & 4.239               & 28.731      & 90.002        & 89.998            &  119.961       \\ 
	Fe(A)-Substituted\tablenotemark[1]      & 4.225      & 4.229               & 28.758      & 89.997        & 89.999            &  119.966       \\
	Fe(B)-Substituted\tablenotemark[1]      & 4.225      & 4.229               & 28.757      & 89.998        & 90.002            &  119.997       \\
		\hline
		\end{tabular}
		\tablenotetext[1]{Variable-cell computational relaxation}
		\label{Table01}
\end{table}

\begin{table}     
	\caption{Bond lengths, atomic  magnetic moments ($|\vec{\mu}|$)\protect\tablenotemark[1]  and
	calculated $^{57}$Fe M\"ossbauer isomer shifts ($\delta_{Fe}$) and
			  electric quadrupole splittings ($\Delta E_{Q}$)
			  for optimized geometries                
			  of two Fe-containing $Bi_2Se_3$ supercells.\protect\tablenotemark[2]                   
                          Experimental~\cite{Pelenovich-2015-TSF} M\"ossbauer parameters for Fe-doped $Bi_2Se_3$  also shown.}
\begin{ruledtabular}
\begin{tabular}{lcc|cc|c|cc}
Supercell  &             &  Site      & Fe-Se(A)       & Fe-Se(B) &  $|\vec{\mu}|$       & $\delta_{Fe}$      & $|\Delta E_{Q}|$       \\
		  &             &                &   [$\AA$]      & [$\AA$]  &  $[\mu_{_B}]$                & [mm/s]             & [mm/s]             \\
	\hline
	(a)	&&Fe(a)              &2.582           &2.773   &3.677   & (+)0.514           &  0.209                      \\
	        &&Se(A)              &                &        &0.126   &                    &                             \\
                &&Se(B)              &                &        &0.051   &                    &                             \\
	(b)	&&Fe(b)              &2.580           &2.775   &3.681   & (+)0.497           &  0.184                      \\
	        &&Se(A)              &                &        &0.128   &                    &                             \\
                &&Se(B)              &                &        &0.051   &                    &                             \\
	Average   & & & &                             &        & (+)0.506           &  0.197                      \\
        (a), (b)  & & & &                             &        &                    &                             \\  
        \hline                                                        
	Exp.~\cite{Pelenovich-2015-TSF}    &&&&          &     & (+)0.44            & 0.23                        \\
	                                   &&&&          &     &  $\pm$0.1          & $\pm$ 0.07                  \\
\end{tabular}
\end{ruledtabular}
\tablenotetext[1]{Magnetic moments in units of Bohr magneton.}    
\tablenotetext[2]{Fe-Se distances correspond to fixed-cell optimizations (Table~\ref{Table01})}                                               
\label{Table02}
\end{table}

\subsection{M\"ossbauer Parameters of $^{57}$Fe Substituents}
Considering their \emph{local} structural environment, Fe$^{3+}$ ions which substitute Bi$^{3+}$ counterparts,
are in a \emph{distorted octahedral} environment. As illustrated by Fig.~\ref{Fig01}(a) such crystalline distortion
gives rise to energy splittings between the iron ion's  3d-electron sub-shells, namely 
t$_{2g}^{X}$ and  e$_g^{Y}$ where X and Y represent the electron occupations of each sub-shell.
To elucidate the origin of spectroscopic parameters one can use     
the \emph{formal} number, $N = X + Y$, of 3d-shell electrons which for the more
common oxidation states Fe$^{2+}$ and Fe$^{3+}$ is  N = 6  and N = 5, respectively.  The 
$^{57}Fe$ M\"ossbauer  parameters of the substituent  atoms were consistent with \emph{nominal}
t$_{2g}^{3}$ - e$_g^{2}$ electronic configurations.
However, non-negligible effects of 3d-shell electron delocalization                
were also present as manifested in the magnitudes of   magnetic moments ($|\vec{\mu}|$) for
iron ions and their selenium-bound neighboors. Table~\ref{Table02} shows that a   \emph{nominal} 
$|\vec{\mu}|$ = 5 $\mu_{_B}$ moment corresponding to  ionic Fe$^{3+}(S=5/2)$ reduces, in a $Bi_2Se_3$ environment, to $\approx$ 3.68$\mu_{_B}$
while the \emph{nominally} diamagnetic $Se^{2-}$ ions acquire  small moments of $\approx$0.13 and $\approx$0.05 $ \mu_{_B}$
for Se(A) amd Se(B), respectively. These results are consistent with the Fe-centered, but Se-delocalized,
spin densities  shown in Figs.~\ref{Fig04}-\ref{Fig05}.

Whereas, depending on the local crystalline fields,  Fe$^{3+}$ ions can be in  high (S = 5/2),
intermediate (S = 3/2) or low (S = 1/2) spin states, their computed electronic configurations and
hyperfine interactions,  as substituents in the $Bi_2Se_3$ host, were  consistent with \emph{nominal}
Fe$^{3+}$(S = 5/2)  states.
This is  reflected in the  predicted  $^{57}$Fe  M\"ossbauer
parameters ($\delta_{Fe} \approx$  0.51 mm/s, $\Delta E_{Q} \approx$  0.20 mm/s, Table~\ref{Table02})
which are, within experimental accuracy,  in excellent  agreement with  experiment                             
for structurally relaxed Fe-containing $Bi_2Se_3$  supercells.

Both parameters, isomer shifts and
quadrupole splittings, are fairly sensitive to the distorted octahedral environments of the iron ions.
The latter parameter,
in particular, reflects the inequivalence of Fe-Se(A) and  Fe-Se(B) bond lengths.
This geometric asymmetry produces minor electric field gradients (EFG) at the iron nuclei and in turn,
\emph{via} Eq.~\ref{eq},  
fairly small electric quadrupole splittings ($|\Delta_{EQ}|$).  Predicted ($\approx$0.20 mm/s) 
and experimental (0.23 mm/s)  values were, within the reported experimental uncertainty ($\pm 0.07$ mm/s)~\cite{Pelenovich-2015-TSF},  in agreement.                          

For other electronic configurations of iron, such
as Fe$^{2+}$(S = 4/2), a more sizable electric
field gradient  generated by 3d-shell electrons  can yield  substantially
larger electric quadrupole splittings.
Thus, in the present case, the presence of $Fe^{2+}$ electronic configurations can be ruled out for  being
inconsistent with the measured M\"ossbauer parameters.
By contrast,  for  Fe$^{3+}$(S = 5/2) ions in distorted octahedral environments,                       
the 3d-shell contributes little to the EFG and its main contribution                   
arises from the crystal lattice. The EFG at the  nuclei of Fe$^{3+}$ substituents in
$Bi_2Se_3$ thin films used for M\"ossbauer measurements,
and consequently their $\Delta E_Q$ values,
mainly originates in the asymmetric position of an iron atom  relative  to  Se$^{2-}$(A) \emph{versus}
Se$^{2-}$(B) ions.

Our findings that iron substituents in the reported M\"ossbauer sample are in a \emph{nominal}  Fe$^{3+}$ state  are 
consistent with   EXD observations~\cite{Li-2013-ML}
that sample annealing suppreses the presence of $Fe^{2+}$ and facilitates substitution of Fe$^{3+}$ for Bi$^{3+}$.        
Likewise, the M\"ossbauer parameters are inconsistent with other possible locations of the substituents, such as
interstitial sites within inter-\emph{quintuple} van der Waal regions, as different values for $\delta_{Fe}$
and/or $\Delta E_Q$ would be expected. Indeed previous  studies suggest  that Fe for Bi substitution
is more energetically favorable than interstitial doping.~\cite{Wei-2015-PLA,Zhang-2012-PRL}

\subsection{Spectroscopic and Structural Correlations}
Being a \emph{local} probe M\"ossbauer spectroscopy measures parameters which
are sensitive to the structural, chemical and electronic properties
of an  Fe ion's \emph{immediate} environment.
One expects shorter or longer  Fe-Se bond lengths,
within a $Bi_2Se_3$ host lattice, to be reflected in the measured values of  $\delta_{Fe}$ and
$\Delta E_{Q}$.
The significantly different Fe-Se bond lengths corresponding to                         
``Substituted Fixed Cell" \emph{versus}  ``Unsubstituted Fixed Cell (Bi$\rightarrow$Fe)" structures, summarized in Table~S1\cite{SuppPRB},
are expected to yield correspondingly  different M\"ossbauer parameters.  Although the Fe substituents in a particular
M\"ossbauer sample may correspond to a specific  geometric environment, computationally it is possible to study  $\delta_{Fe}$ and
$\Delta E_Q$ as a function of geometry. 
Accordingly,  we computed structure-dependent  M\"ossbauer parameters
for  two different Fe ion 
environments corresponding to the dimensions of  energy-minimized   i) ``Substituted Fixed Cell"                          
and  ii) ``Unsubstituted Fixed Cell (Bi$\rightarrow$Fe)"  structures. We recall that  the first type has shorter Fe-Se bond lengths
in comparison to the second type which  has longer bonds.
Since both typed of structures
were obtained by means of the same SPES protocol it is possible to directly compare their
bond lengths and other metric parameters.
As expected, computed M\"ossbauer parameters for both types of structures were substantially different and only those
corresponding to the first type, with  reorganized iron environments as shown in Figs.~\ref{Fig03}(c,d),
were in agreement with  experiment (Table~\ref{Table02}).
For comparison M\"ossbauer results for these and  other structures are listed in Supp. Table~S1\cite{SuppPRB}.

The different $\delta_{Fe}$ and
$\Delta E_Q$ values calculated for different iron environments,  show that 
M\"ossbauer spectroscopy is  sensitive to structural variations.                                          
The discrepancy between  experimental M\"ossbauer values with those                      
corresponding to ``Unsubstituted Fixed Cell (Bi$\rightarrow$Fe)"  geometries,             
in contrast with the excellent  agreement obtained
for those with reorganized Fe environments (``Substituted Fixed Cell"), confirms that the structural and bonding
environments of Fe$^{3+}$ ions in the particular sample used for  M\"ossbauer measurements~\cite{Pelenovich-2015-TSF}   are 
very closely  described by Figs.~\ref{Fig03}(c,d).

\section{Summary}
M\"ossbauer spectroscopy probes the \emph{local} electronic, magnetic and structural environments of
$^{57}$Fe substituents in crystalline environments.
Conversion electron M\"ossbauer measurements were performed on Fe-doped epitaxial thin films by
Pelenovich \emph{et al.}~\cite{Pelenovich-2015-TSF}
However, to establish a direct correlation between  
measured  M\"ossbauer parameters and the geometric structure  of Fe-substituted $Bi_2Se_3$, 
\emph{ab initio} quantum mechanical  analysis was needed.
Accordingly, the present SPES calculations                             
explain the relationship between geometric structure and M\"ossbauer properties of substituted $Bi_2Se_3$ in the limit of low iron concentration.
In this limit, predicted structure-dependent M\"ossbauer parameters were in excellent agreement with experiment.

As Fe$^{3+}$ substites  Bi$^{3+}$ we found a 
substantial, yet \emph{localized}, structural reorganization  which  yields Fe-Se distances 
significantly shorter than  corresponding Bi-Se bonds in the undoped structure.                      
It is within this restructured   environment that substituted
iron yields the experimentally observed M\"ossbauer parameters.
We have shown that
Fe$^{3+}$ ions in a very specific system,
namely the reported epitaxial thin-film M\"ossabuer sample,~\cite{Pelenovich-2015-TSF}
are in a \emph{local} distorted-octahedral Se-rich environment.
Although experimental M\"ossbauer parameters have been reported 
for epitaxial $Bi_2Se_3$ films, essentially the same behavior is expected for
low concentration of Fe doping in  single crystals
if and when Fe$^{3+}$ explicitly substitutes Bi$^{3+}$.
Correlations between geometric, electronic and  magnetic properties  for
Fe-substituted $Bi_2Se_3$ have been established.  An Fe-centered
spin density is reported which extends
towards neighboring chemically-bound Se atoms \emph{via} direct Fe-Se bonding and 
concomitant hybridization of Fe(d) and Se(p) orbitals. The substituted structure,
therefore, has   magnetic moments mainly centered on iron ($\approx$ 3.68 $\mu_B$) 
with small, but finite, contributions from its neighbooring Se(A)  ($\approx$ 0.13 $\mu_B$)
and Se(B) ($\approx$ 0.05 $\mu_B$) atoms.

\vskip 12pt
{\bf Acknowledgement.} Access to high performance computing facilities by Purdue University's ITaP center is gratefully acknowledged.

%

	\end{document}